\documentstyle[12pt,prd,aps,epsfig]{revtex}
\begin{document}
\baselineskip=17pt
\draft
\title{Effects of  conversions for high energy neutrinos originating 
 from cosmological gamma-ray burst fireballs}
\author{H. Athar}
\address{Departamento de F\'\i sica de Part\'\i culas, 
 Universidade de Santiago de Compostela, E-15706 Santiago de Compostela,
Spain; E-mail: athar.husain@cern.ch}
\date{\today}
\maketitle
\begin{abstract} 
\tightenlines

	We study neutrino conversions in the recently envisaged source 
of high energy ($E {\buildrel > \over {_{\sim}}} \, 10^{6}$ GeV)  
neutrinos, that is, in the vicinity of cosmological gamma-ray burst fireballs
 (GRB). We consider mainly 
the possibility of neutrino conversions due to an interplay of neutrino 
 transition magnetic
moment, $\mu$,  and the violation of equivalence principle (VEP), 
 parameterized by $\Delta f$, in
a reasonable strength of magnetic field in the vicinity of the 
 GRB. We point out that for 
 $\Delta f\, \sim 10^{-25}(\delta m^2/1\mbox{eV}^2)$, 
a resonant spin-flavour precession between $\bar{\nu}_{\mu}$ and 
 $\nu_{\tau}$ may occur
in the vicinity of GRB for 
 $\mu \, \sim \, 10^{-12}\, \mu_{B}$ 
 ($\mu_{B}$ is Bohr magneton), 
thus enhancing the expected high energy $\nu_{\tau}$ flux from GRBs.     
\end{abstract}


\section{Introduction}
	Recently, cosmological fireballs are suggested as a 
possible
production site for gamma-ray bursts as well as the high energy 
 ($E {\buildrel > \over {_{\sim}}} \, 10^{6}$ GeV)  neutrinos \cite{WB}. 
Although,
 the origin of these cosmological gamma-ray burst fireballs (GRB) 
 is not 
yet well-understood, the observations \cite{P} suggest that generically 
a very compact source of  linear scale $\sim \, 10^{7}$ cm  through 
 internal or/and external shock propagation produces these 
gamma-ray bursts as well as the (burst of) high energy neutrinos.
 Typically, this compact source is hypothesized to be formed possibly
due to the merging of binary neutron stars or due to collapse of a 
supermassive star. 

The main source of high energy tau neutrinos in GRBs is the production 
 and decay of $D^{\pm}_{S}$. The production of $D^{\pm}_{S}$ may be  
 through $p\gamma $ and/or  through $pp$ collisions.
	In \cite{PX}, the $\nu_{e}$ and $\nu_{\mu}$  flux is 
estimated in $pp$
collisions, whereas in \cite{WB}, the $\nu_{e}$ and $\nu_{\mu}$ flux is 
estimated in $p\gamma$ collisions for GRBs. 
In $pp$ collisions, the flux of $\nu_{\tau}$ may be obtained through the main 
process of
$p+p\rightarrow D^{+}_{S}+X$. The $D^{+}_{S}$ decays into 
$\tau^{+}$ 
lepton and $\nu_{\tau}$ with a branching ratio of $\sim \, 3\%$. This 
$\tau^{+}$ lepton
further decays into $\nu_{\tau}$.
The cross-section for  $D^{+}_{S}$ production, which is main source of 
$\nu_{\tau}$'s,  
is 1-2 orders of magnitude lower than that of $D^{+}$ and/or $D^{-}$ 
 which subsequently produces 
$\nu_{e}$ and $\nu_{\mu}$. The branching ratio for 
$\nu_{e}$ and/or
$\nu_{\mu}$ production is higher upto an order of magnitude 
 than for $\nu_{\tau}$ production (through $D^{\pm}_{S}$). These imply 
that the 
$\nu_{\tau}$ flux in $pp$ collisions is suppressed up to 3-4 orders of 
magnitude 
relative to corresponding $\nu_{e}$ and/or $\nu_{\mu}$ fluxes. 
 In $p\gamma$ collisions, 
the main
process for the production of $\nu_{\tau}$ may be 
  $p+\gamma \rightarrow D^{+}_{S}+
\Lambda^{0}+\bar{D}^{0}$ with similar relevant branching ratios and 
 corresponding cross-section  values. 
 Here the corresponding main source for $\nu_{e}$ and $\nu_{\mu}$ 
 production is 
 $p+\gamma \, \rightarrow \Delta^{+}\rightarrow \pi^{+}+n$. 
Therefore, in $p\gamma $ collisions, the $\nu_{\tau}$ flux is also 
 suppressed up to 3-4 orders 
 of magnitude relative to $\nu_{e}$ and/or $\nu_{\mu}$ flux. Thus, in both 
type of collisions, including the relevant kinematics, the  
$\nu_{\tau}$ flux 
is estimated to be rather small relative to $\nu_{e}$ and/or $\nu_{\mu}$ 
  fluxes from GRBs, typically being, $(\nu_{\tau}+\bar{\nu}_{\tau})
 /(\nu_{\mu}+\bar{\nu}_{\mu})\, 
 {\buildrel < \over {_{\sim}}} \, (10^{-4}-10^{-3})$ \cite{VZA}.

	In this brief report, we consider the possibility of obtaining higher 
$\nu_{\tau}$
neutrino flux, that is, 
 $(\nu_{\tau}+\bar{\nu}_{\tau})/(\nu_{\mu}+\bar{\nu}_{\mu})\, \gg \, 10^{-4}$, 
 from GRBs through neutrino 
conversions
as compared to no conversion situation \cite{athar}. In particular, 
 we obtain the range of neutrino
mixing parameters resulting from an interplay of possible violation of the 
equivalence
principle (VEP) parameterized by $\Delta f$ and the magnetic field in the 
vicinity of GRBs yielding 
 $(\nu_{\tau}+\bar{\nu}_{\tau})/(\nu_{\mu}+\bar{\nu}_{\mu})\, \gg \, 10^{-4}$. 
 The possibility of VEP  arises from the realization 
 that different flavours of neutrinos may 
couple differently to gravity \cite{G}.

	The present study is particularly useful as the new ice/water 
 \v{C}erenkov light neutrino
telescopes namely AMANDA and Baikal (also the NESTOR and ANTARES)
 will have energy, angle 
and flavour resolution for high energy neutrinos originating at cosmological 
distances \cite{M}. Recently, there are several discussions concerning the 
signatures 
of a possible neutrino burst from GRBs correlated in time and 
angle 
\cite{W}. In particular, there is a suggestion of measuring $\nu_{\tau}$ 
flux from 
cosmologically distant  sources through a double bang event \cite{L} or 
 through a small pile up of upgoing $\mu$-like  events near 
 (10$^{4}-10^{5}$) GeV \cite{HS}.

	The plan of this brief report is as follows. 
 In Sect. II, we briefly describe  the matter density
and magnetic field in the vicinity of GRBs. In Sect. III, we discuss in 
some detail, the range of neutrino mixing parameters that may give rise to 
relatively large precession/conversion probabilities resulting from 
 an interplay of VEP 
and the  magnetic field in the vicinity of GRBs for vanishing 
gravity and vacuum mixing angles. 
 In the same Sect., we briefly discuss the relevant neutrino mixing parameter
 range   for non-vanishing
 gravity and vacuum mixing angles with vanishing neutrino magnetic moment.
 In Sect. IV, we give  estimates of separable but contained  double bang
 event rates induced by high energy $\nu_{\tau}$'s originating from GRBs 
  without/with conversions for illustrative purposes  
 and finally in Sect. V, we summarize our results.

\section{The matter density and the magnetic field in the vicinity of GRB}

According to \cite{WB}, 
 the high energy neutrino production may take place in the vicinity of
$r_{p}\, \sim \, \Gamma^{2}c\Delta t$. Here $\Gamma $ is the Lorentz 
factor (typically $\Gamma \, \sim \, 300$) and $\Delta t$ is the observed
GRB variability time scale (typically $\Delta t\, \sim \, 1$ ms). Thus,  
 the fireball matter density is 
$\rho \, \sim \, 10^{-13}$
 g cm$^{-3}$ in the vicinity of $r_{p}$ \cite{WB}. 
 In these models, the typical distance traversed by neutrinos 
may be
taken as, $\Delta r\,  {\buildrel < \over {_{\sim}}}\, (10^{-4}-1)$ pc, 
 in the vicinity of 
 GRB, where 1 pc $\sim \, 3\times 10^{18}$ cm.
 These imply that average effective width 
 of matter traversed by neutrinos originating from GRB  is 
$l_{GRB}\, \simeq \, \rho \times \Delta r\, \sim 10^{4}$ g cm$^{-2}$. 
 In the presence of 
matter, the relevant effective width of matter needed for appreciable  
 spin-flavour conversions is
$l_{\mbox {m}}\, \sim \, \sqrt{2} \pi m_{N}/G_{F}\, \sim \, 
 2\times 10^{9}$ g cm$^{-2}$. 
Thus, $l_{GRB}\, \ll \, l_{\mbox {m}}$, and hence no matter 
 effects are expected due to coherent forward scattering of neutrinos off the
  background for high energy neutrinos originating from GRBs.

Taking the observed duration of the typical gamma-ray burst as, $\Delta t\, 
  {\buildrel < \over {_{\sim}}}
 \, 1$ ms, we obtain the mass of the source as, $M_{GRB}\, 
 {\buildrel < \over {_{\sim}}}\, \Delta t/G_{N}$, where $G_{N}$ is the 
 gravitational constant. 
 Let us mention here that for the relatively shorter observed duration of  
 gamma-ray burst from a typical GRB,
 $\Delta t\, \sim \, 0.2 $ ms, implying $M_{GRB}\, \sim \, 40\, M_{\odot}$
 (where $M_{\odot}\, \sim \, 2\times 10^{33}$ g is solar mass). 
We use $M_{GRB}\, \sim \, 2\times 10^{2}\,M_{\odot}$ in our estimates.  

	The magnetic field in the vicinity of a GRB is obtained 
by considering the equipartition arguments \cite{WB}.
We use the following profile of magnetic field, $B_{GRB}$,  
for $r\, > \, r_{p}$ \cite{mag}
\begin{equation}
 B_{GRB}(r)\, \simeq \, B_{0}\left(\frac{r_{p}}{r}\right)^{2},
\end{equation}
where, $B_{0}\, \sim \, L^{1/2}c^{-1/2}(r_{p}\Gamma)^{-1}$ with $L$ being the 
total wind luminosity (typically $L\, \sim \, 10^{51}$ erg s$^{-1}$).

\section{Neutrino conversions in GRB}

	Consider a system of two mixed neutrinos $\bar{\nu}_{\mu}$ 
and $\nu_{\tau}$ for simplicity. The difference of diagonal elements of 
 the $2\times 2$ effective Hamiltonian describing the dynamics of the 
 mixed system of these 
 oscillating neutrinos in the basis 
 $\psi^{T}\, =\, (\bar{\nu}_{\mu}, \nu_{\tau})$ for vanishing vacuum and 
gravity mixing angles is \cite{AK}
\begin{equation}
 \Delta H\, =\, V_{G}-\delta, 
\end{equation}
whereas, each of the off diagonal elements is $\mu B$ ($\mu $ is 
 neutrino magnetic moment). In Eq. (2), $\delta \, =\, 
\delta m^{2}/2E$, where $\delta m^{2}\, =\, m^{2}
(\nu_{\tau})-m^{2}(\bar{\nu}_{\mu})
\, >\, 0$ and $E$ being the neutrino energy.  
Here $V_{G}$ is the effective potential felt by the neutrinos at a 
distance $r$ 
from a gravitational source of mass $M$ due to VEP and is given by \cite{G}
\begin{equation}
 V_{G} \, = \, \Delta f \phi(r)E,
\end{equation}
where $\Delta f \, =\, f_{3}-f_{2}$ is a measure of the 
degree
of VEP and $\phi(r)\, = \, G_{N}Mr^{-1}$ is the gravitational potential 
in the 
Keplerian approximation. 
 In Eq. (3), $f_{3}G_{N}$ and $f_{2}G_{N}$ are 
 respectively the gravitational couplings of $\nu_{\tau}$ and 
 $\bar{\nu}_{\mu}$, such that $f_{2}\, \neq \, f_{3}$. 
There are three relevant $\phi (r)$'s that need to be considered \cite{MS}.
 The effect of $\phi (r)$ due to supercluster named Great Attractor 
 with $\phi_{SC} (r)$ in the vicinity of GRB;
 $\phi (r)$ due to GRB itself, which is,  
 $\phi_{GRB} (r)$, in the vicinity of GRB and the  
 galactic gravitational potential, which is,  $\phi_{G}(r)$. 
 Therefore, we use, $\phi(r)\, \equiv \, \phi_{SC}(r)+\phi_{GRB}(r)+\phi_{G}(r)$.
However, $\phi_{G}(r)\, \ll \, \phi_{SC}(r),\, \phi_{GRB}(r)$ in the vicinity of GRB.
 Thus, 
$\phi(r)\, \simeq \, \phi_{SC}(r)+\phi_{GRB}(r)$. If the neutrino 
production region  $r_{p}$ is  ${\buildrel < \over {_{\sim}}} 10^{13}$ cm then
at $\sim \, r_{p}$, we have $\phi_{GRB}(r)\, >\, \phi_{SC}(r)$. At $r\, \sim \,
6\times 10^{13}$ cm, $\phi_{GRB}(r)\, \sim \, \phi_{SC}(r)$ and for 
$r\,  {\buildrel > \over {_{\sim}}}\, 10^{14}$ cm, $\phi_{GRB}(r)\, <\,
\phi_{SC}(r)$. If the supercluster is a fake object then $\phi(r)\, \simeq \,
\phi_{GRB}(r)$. 
 Here we assume the smallness of the effect of $\phi (r)$ due to an 
 active galactic nucleus (AGN), if any, nearby to GRB.

The possibility
of vanishing gravity and vacuum mixing angle in Eq. (2) allows us to 
identify the
range of $\Delta f$ relevant for the neutrino magnetic moment effects 
only. Latter in this Sect., we briefly comment on the ranges of relevant 
 neutrino mixing  parameters 
for non-vanishing gravity and vacuum mixing angles with vanishing 
 neutrino magnetic moment.

 The case of $\bar{\nu}_{e}\rightarrow \nu_{\tau}$ can  be studied by
 replacing $\bar{\nu}_{\mu}$ with $\bar{\nu}_{e}$ along with corresponding 
 changes in $V_{G}$ and $\delta m^{2}$. 
Let us here mention that 
initial flux of $\bar{\nu}_{e}$ may be smaller than that of $\nu_{\mu}$ by 
a factor of less than 2 \cite{WB}, thus also possibly enhancing the expected 
 $\nu_{\tau}$ flux from GRBs through $\bar{\nu}_{e}\rightarrow \nu_{\tau}$. 
 However, we have checked that observationally 
 this possibility leads to quite similar results in terms of 
 event rates and are therefore not
 discussed here further. We now intend to study in 
some detail, the various possibilities arising from relative 
comparison between $\delta $ and $V_{G}$ in Eq. (2).

	Let us first ignore the effects of VEP ($\Delta f\, =\, 0$). For 
constant $B$, the spin-flavour precession probability $P(\bar{\nu}_{\mu}\rightarrow 
 \nu_{\tau})$ is obtained using Eq. (2) as
\begin{equation}
 P(\bar{\nu}_{\mu}\rightarrow \nu_{\tau})\, =\, 
 \left[\frac{(2\mu B)^{2}}{(2\mu B)^{2}+\delta^{2}}\right]
 \sin^{2}\left(\sqrt{(2\mu B)^{2}+\delta^{2}}
 \cdot \frac{\Delta r}{2}\right). 
\end{equation}
We take $\mu \, \sim \, 10^{-12}\, \mu_{B}$
or less, where $\mu_{B}$ is Bohr magneton, which is less than
the stringent astrophysical upper bound on $\mu$ based on cooling of 
red giants \cite{R}. 
We here consider the transition magnetic moment,  
allowing the
possibility of simultaneously changing the neutrino flavour as well as the helicity 
. Thus,  the precessed  
$\nu_{\tau}$ is an active neutrino and interacts weakly. 
 If $\mu $ is of Dirac type, then the precession leads to 
 disappearance/nonobservation of $\nu_{\tau}$
 as the precessed $\nu_{\tau}$ is now a sterile one. In Eq. (4), $\Delta r$
 is the width of the region with $B$. Note that if 
$\delta \, \ll \, 2\mu B$,
 then, for $E\, \sim \, 2\times 10^{6}$ GeV and using Eq. (1), 
 we obtain $\delta m^{2}\, \ll \, 
 5\times 10^{-8}$ eV$^{2}$. We take, $\delta m^{2}\, \sim \, 10^{-9}$
  eV$^{2}$, as an example and consequently we obtain from Eq. (4) an energy 
 independent large ($P\, >\, 1/2$) spin-flavour 
precession probability for $\mu \, \sim \, 10^{-13}\mu_{B}$
 with $10^{-4}\, {\buildrel < \over {_{\sim}}}\, \Delta r/\mbox{pc}
 {\buildrel < \over {_{\sim}}}\,1$ . 
 Let us mention here that the typical relevant energy span for 
 high energy neutrinos originating from the considered class of GRBs is an
  order of magnitude, that is, $2\times 10^{6} {\buildrel < 
\over {_{\sim}}}\, E/\mbox{GeV} \, {\buildrel < \over {_{\sim}}}\, 2\times 10^{7}$
 (see Sect. IV). Thus, for $\mu $ of the 
 order of $10^{-13}\, \mu_{B}$,
the $\nu_{\tau}$ flux may be higher than the expected one from GRBs
 in the absence of spin-flavour precessions, that is,
 $(\nu_{\tau}+\bar{\nu}_{\tau})/(\nu_{\mu}+\bar{\nu}_{\mu})\, \gg \, 10^{-3}$
 due to neutrino spin-flavour precession effects. 
 Since the neutrino spin-flavour precession effect is
 essentially determined by the product $\mu B$ so one may rescale  $\mu $ and $B$
 to obtain the same results. 
 For $\delta \, \simeq 2\mu B$ and $\delta \, \gg \, 
 2 \mu B$, we obtain from Eq. (4), an energy dependent $P$ such
 that $P\, <\, 1/2$.

	With non-vanishing $\Delta f$ ($\Delta f\, \neq \, 0$), a resonant 
character 
in neutrino spin-flavour precession can be obtained for a range of  
 values of relevant neutrino 
mixing parameters \footnote{\footnotesize\tightenlines From above discussion, 
 it follows that $E$ 
 dependent/independent spin-flavour precession  
   may also be obtained for nonzero $\Delta f$, 
 however, given the current status of the high energy neutrino detection, 
 for simplicity, we ignore these possibilities which tend to 
 overlap with this case for a certain range of relevant 
 neutrino mixing parameters; for details of these possibilities in the 
 context of  AGN, see \cite{athar}.} Two conditions are 
 essential to obtain a resonant character in
neutrino spin-flavour precession: the level crossing and the adiabaticity 
condition.
The level crossing condition is obtained by taking $\Delta H\, =\, 0$ and 
 is given by:
\begin{equation}
 \delta m^{2}\, \sim \, \left(\frac{|\Delta f|}{10^{-25}}\right)\, \mbox{eV}^{2}.
\end{equation}  
The other essential condition, namely, 
 the adiabaticity in the resonance  reads \cite{AK}
\begin{equation}
 \kappa \, =\, \frac{2(2\mu B)^{2}}{|\mbox{d}V_{G}/\mbox{d}r|}.
\end{equation}
Note that here  $\kappa $ depends
explicitly on $E$ through $V_{G}$ unlike the case of ordinary neutrino 
 spin-flip induced by the matter effects. 
 A resonant character in neutrino spin-flavour precession is obtained if 
$\kappa \,  {\buildrel > \over {_{\sim}}}\, 1$ such that Eq. (5) is satisfied. 
 We notice that, 
$B_{\mbox {ad}}/B_{GRB}\, {\buildrel <\over {_{\sim}}}\, 1$ for $\mu \, 
 \sim 10^{-12}\mu_{B}$. Here $B_{\mbox{ad}}$ is obtained by setting 
$\kappa \, \sim \, 1$ in Eq. (6). 
 The general expression for relevant neutrino spin-flavour conversion 
 probability is given by \cite{ICTP} 
\begin{equation}
 P(\bar{\nu}_{\mu}\rightarrow \nu_{\tau})\, =\, 
 \frac{1}{2}-\left(\frac{1}{2}-P_{LZ}\right)\cos 2\theta_{B},
\end{equation}
where $P_{LZ}\, =\, \exp(-\frac{\pi}{2}\kappa)$ and $
 \tan 2\theta_{B}\, =\, (2\mu B)/\Delta H$
 is being evaluated at the high energy neutrino 
 production cite in the vicinity of GRB. 
In Fig. 2, we plot $P(\bar{\nu}_{\mu}\rightarrow \nu_{\tau})$, 
 given by Eq. (7), as a function of neutrino energy $E$ (GeV) for different
$\Delta f $ as well as $\delta m^{2}$ values. 
 Note that the resonant spin-flavour precession probability is $>$ 1/2 for a
 relatively large range of $\delta m^{2}$ (and $\Delta f$ values).  
The expected   spectrum $F(\nu_{\tau}+\bar{\nu}_{\tau})$ of the high energy tau 
 neutrinos originating from GRBs due to spin-flavour conversions 
 is calculated as \cite{ICTP}
\begin{equation}
 F(\nu_{\tau}+\bar{\nu}_{\tau})=[1-
 P(\bar{\nu}_{\mu}\rightarrow \nu_{\tau})]F^{0}(\nu_{\tau}+\bar{\nu}_{\tau})+
 P(\bar{\nu}_{\mu}\rightarrow \nu_{\tau})F^{0}(\nu_{\mu}+\bar{\nu}_{\mu}),
\end{equation}
where $F^{0}$'s are the neutrino flux spectrums originating from GRBs. 
In Fig. 3, we plot the expected $\nu_{\tau}$ spectrum obtained  by 
neutrino spin-flavour precessions and conversions (induced by an interplay 
 of VEP and the magnetic field 
in GRBs). We use the GRB spectrum for $(\nu_{\mu}+\bar{\nu}_{\mu})$, i.e., 
 $F^{0}(\nu_{\mu}+\bar{\nu}_{\mu})$ from
\cite{WB} and $F^{0}(\nu_{\tau}+\bar{\nu}_{\tau})$ from \cite{VZA} and 
 multiply these  by respective $P(\bar{\nu}_{\mu}\rightarrow \nu_{\tau})$ 
 given by Eq. (8) to obtain $F(\nu_{\tau}+\bar{\nu}_{\tau})$ due to 
 resonant spin-flavour precession
 (lower curve in Fig. 2). The upper curve in Fig. 2 is obtained by multiplying 
 $P$ given by Eq. (4) for small $\delta m^{2}$. 

	Let us now consider briefly the effects of nonvanishing gravity 
 mixing angle  
$\theta_{G}$ and non-vanishing vacuum mixing angle $\theta$ for vanishing 
 neutrino magnetic moment. 
	If $\theta_{G}\, =\, 0$ and $\theta\, \neq 0$, then taking the 
 distance 
between a typical GRB and our galaxy as, $L\, \sim \, 1000 $ Mpc, the vacuum 
 flavour oscillations may occur between $\nu_{\mu}$ and $\nu_{\tau}$ for  
 $\delta m^{2}\, \sim \, 10^{-3} \mbox{eV}^{2}$
 with maximal vacuum flavour mixing between $\nu_{\mu}$ and $\nu_{\tau}$. 
 These values of neutrino mixings have been suggested as a  possible 
 explanation of recent superkamiokande data 
 concerning the deficit of atmospheric muon neutrinos \cite{skk}.
 The corresponding expression for flavour oscillations in vacuum is
\begin{equation}
 P(\nu_{\mu}\rightarrow \nu_{\tau})=\sin^{2}2\theta  \sin^{2}
 \left(\frac{\delta m^{2}}{4E}L\right).
\end{equation}
Note that here the resulting $P$ is $\sim \, 1/2$ due to the fact that 
 $4E/\delta m^2 \, \ll \, L$.  
For $\theta_{G}\, \neq \, 0$ and $\theta\, =\, 0$, in the case of massless or
degenerate neutrinos, the corresponding vacuum flavour oscillation analog 
 for $\nu_{e}\rightarrow \nu_{\tau}$ is obtained through 
 $\theta \rightarrow \theta_{G}$ and $\frac{\delta m^{2}}{4E} \rightarrow 
 V_{G}$. 
For maximal $\theta_{G}$, the sensitivity of $\Delta f$
 may be estimated by equating the argument of second $\sin$ factor equal to
$\pi/2$ in the corresponding expression for $P$ \cite{MS}. 
 This implies $\Delta f\, \sim \, 10^{-41}$ with $\phi(r)\, \simeq \, 
 \phi_{SC}(r)$. Note that this value of $\Delta f$ is
 of the same order of magnitude as that expected for neutrinos originating 
 from AGNs \cite{MS}. In contrast to neutrino spin-flavour precession effects
 given by Eq. (4) resulting in $P\, >\, 1/2$, the vacuum flavour oscillations 
 give $P\, \sim \, 1/2$, thus, allowing the possibility of 
 isolating the mechanism
 of oscillation since the high energy neutrino telescopes may attempt to measure
 $(\nu_{\tau}+\bar{\nu}_{\tau})/(\nu_{\mu}+\bar{\nu}_{\mu})$. 
 Further, with the improved information on either $\Delta f$
 and/or $\mu $, one may be able to  distinguish between the situations of
 resonant and nonresonant spin-flavour precession  induced by  an interplay 
 of VEP and $\mu $ in $B_{GRB}$. Concerning possibilities of resonant
 flavour conversion, a resonant or/and nonresonant flavour conversion 
 between $\nu_{e}$ and $\nu_{\tau}$ 
 in the vicinity of a GRB is also possible due to an
 interplay of vanishing/nonvanishing vacuum and gravity mixing angles 
 \cite{MS}. For instance, with $\theta \, \rightarrow \, 0$, 
 a resonant flavour conversion between $\nu_{e}$ and $\nu_{\tau}$ may be 
 obtained if $\sin^{2}2\theta_{G}\, \gg \, 0.25$. Here the relevant level 
crossing may occur at $r\, \sim \, 0.1 $ pc with 
 $\delta m^{2}\, \sim \, 10^{-6}$ eV$^{2}$. 

\section{Signatures of high energy $\nu_{\tau}$ in neutrino telescopes}

	The  km$^{2}$ scale high energy neutrino telescopes may 
 be able to obtain first examples of high energy $\nu_{\tau}$, 
 through {\em double bangs}, 
originating from GRBs correlated in time and direction with corresponding 
gamma-ray burst \cite{L}. 
 The first bang occurs due to charged current interaction of high energy 
 tau neutrinos near/inside the neutrino telescope producing the tau lepton 
 and the second bang occurs due to hadronic decay of this tau lepton. 
Following \cite{APZ}, we present in Table I, the expected 
 contained but separable  double bang event rates for downgoing $\nu_{\tau}$ 
in 1 km$^{2}$ size water \v{C}erenkov high energy neutrino telescopes 
 for illustrative purposes. 
To calculate the event rates, we use Martin Roberts Stirling 
 (MRS 96 R$_{1}$) parton distributions \cite{mrs} and present event rates 
 in units of yr$^{-1}$
sr$^{-1}$. We have checked that other recent parton  distributions give quite 
 similar event rates and are therefore not depicted here.   
From Table I, we notice that the event 
 rates for neutrino spin-flavour precession are up to $\sim $ 4 orders of 
 magnitude higher than that for typical intrinsic (no oscillations) tau   
 neutrino flux.

The condition of containedness is obtained by requiring that the 
 separation between the two bangs is less than the typical $\sim $ km 
size of the neutrino telescope. It is obtained by equating the range of tau 
 neutrino induced tau leptons with the size of detector implying 
$E \,  {\buildrel < \over {_{\sim}}}\, 2\times 10^{7}$ GeV. 
 The condition of separableness is 
obtained by demanding the separation between the two bangs is larger than 
 the typical spread of the bangs such that the amplitude of the second bang
 is essentially 2 times the first bang. This leads to 
$E \,  {\buildrel > \over {_{\sim}}}\, 2\times 10^{6}$ GeV \cite{L}. 
 Thus, the two bangs may be separated by a $\mu$-like track within these energy
 limits. The upgoing 
 tau neutrinos at these energies may lead to a small pile up of upgoing $\mu$-like 
 events near (10$^{4}-10^{5}$) GeV with flat zenith angle dependence \cite{HS}.

The possibility of measuring the contained but separable double bang events
 may enable one to distinguish between the high energy tau neutrinos and 
 electron
 and/or muon neutrinos originating from cosmologically distant GRBs 
 while providing useful information about 
 the relevant energy interval at the same time. 
 The chance of having  double bang events
 induced by electron and/or muon neutrinos is negligibly small
 for the relevant energies \cite{L}.

\section{Results and discussion}

	We have studied in some detail the effects of neutrino spin-flavour 
 conversions 
 due to an interplay of the effect of possible VEP  
  and the magnetic field in GRBs  and have 
 obtained the relevant range of neutrino
 mixing parameters for appreciable neutrino conversion probabilities. We have
also briefly commented on the corresponding range of neutrino mixing 
 parameters 
 for vanishing neutrino magnetic moment. 

	The matter density in the vicinity of GRB
is quite small (upto 4-5 orders of magnitude) to induce any resonant 
 flavour/spin-flavour neutrino 
 conversion due to normal matter effects.
We have pointed out that a resonant character in the neutrino spin-flavour 
 conversions 
may nevertheless be obtained due to possible VEP. The 
corresponding degree of VEP  may be
 $\sim \, (10^{-35}-10^{-25}$) depending on $\delta m^{2}$ value.

	The double bang event rate for intrinsic 
 (no oscillations/conversions) high energy
 tau neutrinos originating from GRBs turns out to be small as compared to 
 that due to precession/conversion effects up to a factor of 
 $\sim \, 10^{-4}$. Thus, the high energy neutrino telescopes may
 provide useful upper bounds on intrinsic properties of neutrinos such as
 mass, mixing and magnetic moment, etc.. The relevant tau neutrino energy 
 range for detection in 1 km$^{2}$ neutrino telescopes may be  
 $2\times 10^{6}
 {\buildrel < \over {_{\sim}}}\, E/\mbox {GeV}\,
{\buildrel < \over {_{\sim}}} \, 2\times 10^{7}$ through characteristic 
 contained but separable double bang events. 

Observationally, the high energy $\nu_{\tau}$ burst from a GRB may possibly  
be correlated to the corresponding gamma-ray burst/highest energy cosmic 
 rays (if both have common origin) in time and in 
direction thus raising the possibility of its detection even if there is 
  relatively 
 large background flux of high energy tau neutrinos from AGNs also from 
  oscillations. If the range of neutrino mixing parameters 
 pointed out in this 
study is realized terrestially/extraterrestially then a relatively large 
$\nu_{\tau}$ flux from GRBs is 
expected as compared to no oscillation/conversion scenario.

Since the high energy neutrino telescopes may measure the ratio 
 of the sum of tau and antitau neutrinos and the muon and antimuon 
 neutrinos, so in principle, any change in the expected 
 relatively small (upto 4 orders of magnitude) ratio may be 
 attributed to spin-flavour precession/conversion effects as the 
 $\nu_{\mu}\, \rightarrow \, \nu_{s}$ channel may alternatively be a 
 possible explanation of recent superkamiokande results concerning  the 
 deficit of atmospheric muon neutrinos. 

The flavour/spin-flavour conversions may occur possibly through several
 mechanisms. We have discussed in some detail mainly the spin-flavour
 precession/conversion situation induced by a nonzero neutrino magnetic moment
 and by a relatively small violation of equivalence principle as an example  
 to point out the possibility of obtaining higher tau neutrino 
 fluxes as compared
 to no oscillations/conversions scenarios from gamma-ray burst fireballs.

\paragraph*{Acknowledgments.} 

The author thanks Alexei Smirnov and Enrique Zas for useful discussion and 
 Agencia Espa\~nola de Cooperaci\'on Internacional 
 (AECI), Xunta de Galicia
(XUGA-20602B98) and CICYT (AEN96-1773) for financial support.

\pagebreak

\begin{table}[th]
\caption{Event rate (yr$^{-1}$sr$^{-1}$) for high energy tau neutrino 
 induced contained but separable double
 bangs in various energy bins using MRS 96 parton distributions.}
\begin{tabular}{cccc}  
 {\em Energy Interval} & 
 \multicolumn{3}{c}{\em Rate (yr$^{-1}$sr$^{-1}$)} \\ \cline{2-4}
    & {\em no  osc} & {\em spin-flavour precession} 
 & {\em vac osc} \\ \cline{1-4}
 $2\times 10^{6}{\buildrel < \over {_{\sim}}}\, E/\mbox {GeV}\,
{\buildrel < \over {_{\sim}}} \, 5\times 10^{6}$  &$10^{-5}$ &
 $1\times 10^{-1}$ & $0.5\times 10^{-1}$   \\ 
 $5\times 10^{6}{\buildrel < \over {_{\sim}}}\, E/\mbox {GeV}\,
{\buildrel < \over {_{\sim}}} \, 7\times 10^{6}$  &  $2\times 10^{-6}$ &
 $2\times 10^{-2}$  & $10^{-2}$  \\ 
  $7\times 10^{6}{\buildrel < \over {_{\sim}}}\, E/\mbox {GeV}\,
{\buildrel < \over {_{\sim}}} \, 1\times 10^{7}$   & $2\times 10^{-6}$ &
 $2\times 10^{-2}$   & $10^{-2}$ \\
  $1\times 10^{7}{\buildrel < \over {_{\sim}}}\, E/\mbox {GeV}\,
{\buildrel < \over {_{\sim}}} \, 2\times 10^{7}$   & $2\times 10^{-6}$ & 
 $2\times 10^{-2}$   & $10^{-2}$ 
\end{tabular}
\end{table}

\pagebreak

\begin{figure}[!ht]
\tightenlines
\centerline{\psfig{figure=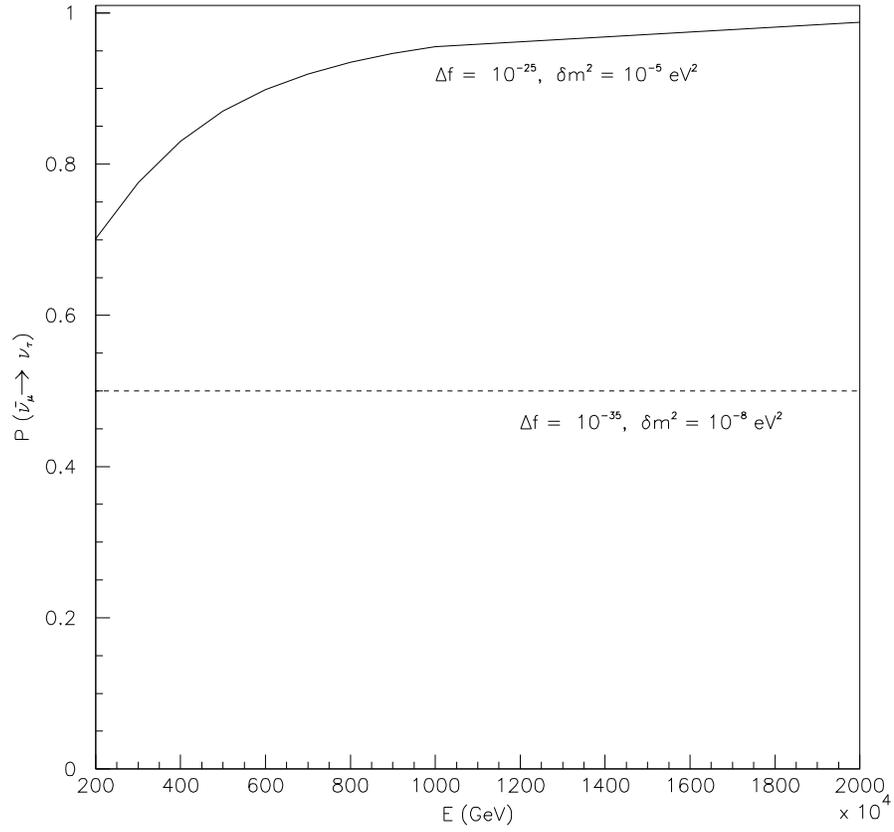,width=5.in}}
\caption[]{
  $P(\bar{\nu}_{\mu}\rightarrow \nu_{\tau})$ 
 using Eq. (7) as a 
 function of $E$ (GeV) for different $\delta m^{2}$ 
 and $\Delta f$ values.
}
\end{figure}  

\pagebreak

\begin{figure}[!ht]
\tightenlines
\centerline{\psfig{figure=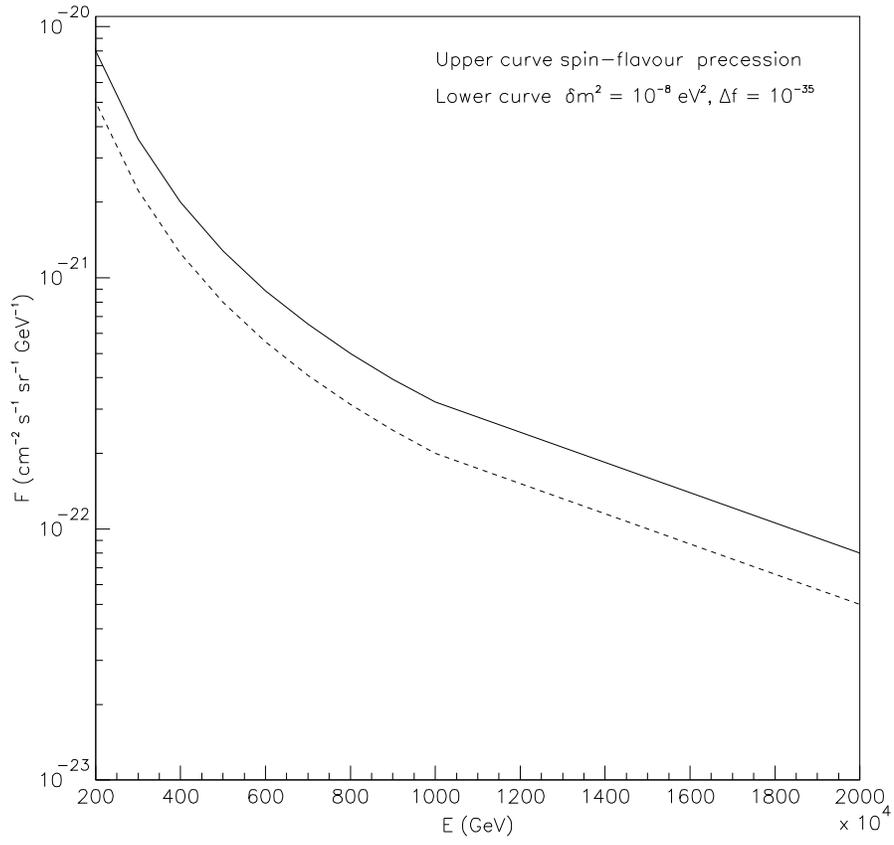,width=5.in}}
\caption[]{
Expected $\nu_{\tau}$ flux spectrum due to different 
 spin-flavour conversion mechanisms.
}
\end{figure}

\end{document}